\documentclass{ws-procs9x6}

\begin{document}

\begin{flushright}
IFUP-TH/2008-34
\end{flushright}

\title{$SO$ AND $USp$ (HYPER)K\"AHLER QUOTIENTS AND LUMPS} 

\author{SVEN BJARKE GUDNASON}

\address{Department of Physics, University of Pisa,\\
INFN, Sezione di Pisa, \\
Largo Pontecorvo, 3, Ed. C, 56127 Pisa, Italy\\
E-mail: gudnason(at)df.unipi.it}

\begin{abstract}
The properties of supersymmetric gauge theories in the Higgs phase at
low energies can appropriately be studied by means of a non-linear
$\sigma$ model, which has the target space being K\"ahler for ${\cal
  N}=1$ supersymmetric models and hyperK\"ahler for ${\cal N}=2$
models. By construction of the K\"ahler and hyperK\"ahler quotients
for the gauge theories with $SO$ and $USp$ gauge symmetries, we obtain
the explicit metrics on their respective manifolds. 
Furthermore, we study the lumps in the non-linear $\sigma$ 
models and their effective description, using the K\"ahler quotients.
\end{abstract}


\bodymatter

\section{Introduction}

The target space of the ${\cal N}=2$ supersymmetric 
non-linear sigma model (NL$\sigma$M), with eight supercharges,
must be hyperK\"ahler \cite{AlvarezGaume:1981hm}. 
By using this fact, the notion of the hyperK\"ahler quotient was
first found in physics  
\cite{Curtright:1979yz,Lindstrom:1983rt} 
and was later formulated mathematically
\cite{Hitchin:1986ea}. 
A $U(1)$ hyperK\"ahler quotient \cite{Curtright:1979yz} 
recovers the Calabi metric on the cotangent bundle 
over the projective space, $T^{\star} {\mathbb C}P^{N-1}$, 
while its $U(N)$ generalization 
leads to the cotangent bundle 
over the complex Grassmann manifold, $T^{\star} G_{M,N}$ 
\cite{Lindstrom:1983rt}. 
The hyperK\"ahler quotient offers a powerful tool to 
construct hyperK\"ahler manifolds such as 
instanton moduli spaces 
\cite{Atiyah:1978ri}, gravitational instantons 
\cite{Kronheimer:1989zs} 
and monopole moduli spaces \cite{Gibbons:1996nt}.

The Higgs branch of  
${\cal N}=2$ supersymmetric QCD is hyperK\"ahler.
The low energy effective theory on the Higgs branch  
is described by an ${\cal N}=2$ NL$\sigma$M 
on the hyperK\"ahler manifold
\cite{Argyres:1996hc,Seiberg:1994aj,Argyres:1996eh}.
In the cases of an $SU(N)$ or a $U(N)$ gauge theory with
hypermultiplets charged commonly under $U(1)$, 
the explicit metrics on the Higgs branch 
and their K\"ahler potentials are known explicitly \cite{}. 
The latter is nothing but the Lindstr\"om-Ro\v{c}ek metric 
\cite{Lindstrom:1983rt,Antoniadis:1996ra}. 
A $U(1) \times U(1)$ gauge theory for instance
gives the space: $T^{\star} F_n$ with 
$F_n$ being the Hirzebruch surface \cite{Eto:2005wf}.

This contribution has two main concerns, the first is the
construction of the metric and K\"ahler potential on the Higgs branch
of ${\cal N}=2$ supersymmetric gauge theories with $SO(2N),\ USp(2N),\
U(1)\times SO(2N),\ U(1)\times USp(2N)$ gauge 
groups\cite{Eto:2008qw}.

The second motive is concerned with $\sigma$ model lumps, or $\sigma$
model instantons. 
A lump solution was first found in the $O(3)$ sigma model, or the
${\mathbb C}P^1$ model \cite{Polyakov:1975yp}. 
It was then generalized to the ${\mathbb C}P^n$ model and the
Grassmann model.  
Gauge theories coupled to several Higgs fields often admit semi-local
vortex-strings \cite{Vachaspati:1991dz}.   
In the strong gauge coupling limit, the gauge theories reduce to
NL$\sigma$Ms and in this limit, semi-local strings reduce to
lump-strings.   
In the gauge theories at finite couplings, the large distance behavior
of semi-local strings is well approximated by lump solutions.  
It was demonstrated in Ref.~\refcite{Eto:2007yv} that non-Abelian
semi-local strings\cite{Shifman:2006kd,Hanany:2003hp} in a $U(N)$
gauge theory reduce to the Grassmann lumps at large distance.  

This work has been done in collaboration with M.~Eto, T.~Fujimori,
K.~Konishi, T.~Nagashima, M.~Nitta, K.~Ohashi and W.~Vinci. Many
details are left in the Ref.~\refcite{Eto:2008qw}.

\section{Obtaining the Low Energy Effective Theory}

Obtaining the low-energy effective theory of supersymmetric gauge
theories on the Higgs branch has been well studied in the literature. 

To obtain the target space, we can do one of the
following\cite{Luty:1995sd}
\begin{romanlist}
\item Fix the gauge to the Wess-Zumino gauge and find the
  potential zeroes ($D=0,\ F=0$) and then mod out the remaining gauge
  group.
\label{sbg:method1}
\item Take the infinite gauge coupling limit immediately and then mod
  out the full complexified gauge group.
\label{sbg:method2}
\item Construct all gauge invariants and find all relations between
  them. This set constitutes the target space. 
\label{sbg:method3}
\end{romanlist}

In the next Section we will construct the metrics following the
method (\ref{sbg:method2}) and rewrite them according to method
(\ref{sbg:method3}) for the gauge theories with $SO({N_{\rm c}})$ and
$USp(2{M_{\rm c}})$ gauge groups. Similarly, the result can be
transformed onto the form of method (\ref{sbg:method1}).  

\section{The $SO({N_{\rm c}})$ and $USp(2{M_{\rm c}})$ K\"ahler Quotients}
\label{sbg:sec:kahler_so_usp}

The K\"ahler potential for an $SO({N_{\rm c}})$ or a $USp(2{M_{\rm
    c}})$ gauge theory is given by
\begin{eqnarray}
 K_{SO,USp} = {\rm Tr}\left[QQ^\dagger e^{-V'}\right] \ , 
 \label{sbg:eq:Kahler_SOUSp}
\end{eqnarray}
where $V'$ takes a value in the $\mathfrak{so}({N_{\rm c}})$ or
$\mathfrak{usp}(2{M_{\rm c}})$ algebra and hence satisfies 
\begin{eqnarray} 
V'^{\rm T}J+JV'=0\quad  \leftrightarrow
\quad e^{-V'{}^{\rm T}} J e^{-V'} = J\ . \label{sbg:eq:V'}
\end{eqnarray} 
Note that, this condition implies that
$\det(e^{-V'})=1$.\footnote{
In the $SO({N_{\rm c}})$ cases, we remove an integral region with
$\det e^{-V}=-1$ in the functional integral of $V$.}
Here the matrix $J$ is the invariant tensor of the
$SO$ or $USp$ group, $g^{\rm T}J g=J$ with $g \in SO({N_{\rm
    c}}),USp(2{M_{\rm c}})$, 
satisfying 
\begin{eqnarray}
J^{\rm T} = \epsilon J\ ,\quad
J^\dagger J =  {\bf 1}_{{N_{\rm c}}}\ ,\qquad
\epsilon = 
\left\{
\begin{array}{cl}
+1 & \text{for}\quad SO({N_{\rm c}})\ ,\\
-1 & \text{for}\quad USp({N_{\rm c}}=2{M_{\rm c}})\ .
\end{array}
\right. \nonumber
\end{eqnarray}
Conversely, a matrix $J$ satisfying the above equations defines a
representation of the $SO$ and $USp$ groups.
We will (mainly) use the convention (in the even case)
\begin{eqnarray}
J \equiv  
\left(
\begin{array}{cc}
{\bf 0}_{M_{\rm c}} & {\bf 1}_{M_{\rm c}}\\
\epsilon {\bf 1}_{M_{\rm c}} & {\bf 0}_{M_{\rm c}}
\end{array}
\right). \nonumber
\end{eqnarray}

First we will discuss the breaking pattern of the gauge symmetry and
the flat directions of the vacuum. 
For this we will consider both the gauge and the global symmetries. 
The vacuum expectation value of $Q_{\rm wz}^{SO}$ in the case of 
$SO({N_{\rm c}})$ can be put on diagonal form after fixing both the local and
the global symmetry\cite{Intriligator:1995id}
\begin{align}
Q_{\rm wz}^{SO} &= \left(A_{{N_{\rm c}}\times {N_{\rm c}}},{\bf 0}_{{N_{\rm c}}\times
  ({N_{\rm f}}-{N_{\rm c}})}\right)\ , \nonumber\\ 
{\rm with} \quad 
A_{{N_{\rm c}}\times {N_{\rm c}}} &= {\rm diag}(a_1,a_2,\cdots,a_{{N_{\rm c}}}) \ , \quad
J={\bf 1}_{{N_{\rm c}}} \ , \nonumber
\end{align}
where all the parameters $a_i$ are taken to be real and positive, which
indeed parametrize flat directions of the Higgs branch. 
In generic points on the vacuum manifold with non-degenerate $a_i$,
there is no global symmetry in the vacuum and the flavor symmetry is
$U({N_{\rm f}})$ apart from $U({N_{\rm f}}-{N_{\rm c}})$ which freely
acts on the vacuum configuration. At a generic point, the vacuum
manifold can be written as  
\begin{eqnarray}
{\cal M}_{SO(N_{\rm c})}^{\rm generic, K} \simeq \mathbb{R}_{\geq
  0}^{{N_{\rm c}}}\times 
\frac{U({N_{\rm f}})}{U({N_{\rm f}}-{N_{\rm c}})}\ .
\label{sbg:eq:SOvacuum_manifold}
\end{eqnarray}
The flat directions ${\mathbb R}_{\geq 0}^{{N_{\rm c}}}$ have no
origin due to symmetry breaking, and their coordinates $\{a_i\}$ are
quasi Nambu-Goldstone (NG) modes \cite{Bando:1983ab}. 
When two parameters coincide, $a_i=a_j, (i\not =j)$,  
a color-flavor locking symmetry $SO(2)$ emerges.
In such degenerate points on the manifold, the above quotient space
attached to ${\mathbb R}_{\geq 0}^{{N_{\rm c}}}$ shrinks to a space
with a smaller dimension.  
The existence of quasi-NG modes are strongly related to the emergence 
of an unbroken phase (classically).
When any two of the $a_i$'s vanish, an $SO(2)$ subgroup of the gauge
symmetry remains unbroken. 
Thus, in the Higgs phase with completely broken gauge symmetry, 
the rank of $Q_{\rm wz}$ has to be greater than ${N_{\rm c}}-2$. 
We will only consider this latter case here, namely the models with
${N_{\rm f}} \ge {N_{\rm c}}-1$. 

For the $USp(2{M_{\rm c}})$ case \cite{Intriligator:1995ne,Argyres:1996hc}, 
it is known that the flat directions are parametrized by
\begin{eqnarray} 
Q_{\rm wz}^{USp} = \left(A_{{M_{\rm c}}\times {M_{\rm c}}}, {\bf
  0}_{{{M_{\rm c}}}\times ({M_{\rm f}}-{M_{\rm c}})}\right) \otimes {\bf 1}_2 \ , 
\nonumber
\end{eqnarray}
where the number of flavors is even ${N_{\rm f}} = 2{M_{\rm f}}$.
Even in generic points with non-degenerate $\{a_i\}$,
the color-flavor symmetry $USp(2)^{{M_{\rm c}}}\simeq SU(2)^{{M_{\rm c}}}$ remains
unbroken in the vacuum. 
Therefore, the vacuum manifold can, at generic points, be written as 
\begin{eqnarray}
{\cal M}_{USp(2M_{\rm c})}^{\rm generic, K}\simeq {\mathbb R}_{\geq
   0}^{{M_{\rm c}}}\times \frac{U({N_{\rm f}})}{SU(2)^{{M_{\rm
         c}}}\times U({N_{\rm f}}-{2M_{\rm c}})}\ , 
\label{sbg:eq:USpvacuum_manifold}
\end{eqnarray}
except for sub-manifolds where the quotient space shrinks.
In this case the completely broken gauge symmetry needs ${M_{\rm f}} \ge
{M_{\rm c}}$.

The $D$-flatness conditions in the Wess-Zumino gauge (i.e.~method
(\ref{sbg:method1})) are 
\begin{eqnarray}
 D^A = {\rm Tr}_{\rm f} \left(Q_{\rm wz}^\dagger T^A Q_{\rm wz}\right) = 0\ , 
\nonumber
\end{eqnarray}
with $T_A$ being the generators in the Lie algebra $\mathfrak{so}$ or
$\mathfrak{usp}$. However, these conditions are rather difficult to
solve.\footnote{To our knowledge the $D$-flatness conditions have  
not been solved in the case of an $SO$ or a $USp$, ${\cal N}=1$ 
supersymmetric gauge theory. }
Without taking the Wess-Zumino gauge, we can eliminate the superfield
$V'$ directly within the superfield formalism by using the following trick; 
let us consider $V'$ taking a value in a larger algebra, namely
$\mathfrak{u}({N_{\rm c}})$ and then introduce an ${N_{\rm c}}\times{N_{\rm c}}$
matrix of Lagrange multipliers\footnote{
Hermiticity of $\lambda$ is defined 
such that $\lambda e^{-V'^{\rm T}}J$ is a vector
superfield, that is, $\lambda^\dagger=e^{V'^{\rm T}}J \lambda\,
e^{-V'^{\rm T}}J$.}
$\lambda$ to restrict $V'$ to take a value in the
$\mathfrak{so}({N_{\rm c}})$ or the 
$\mathfrak{usp}({N_{\rm c}}=2{M_{\rm c}})$ sub-algebra which leaves us
with 
\begin{eqnarray}
 K_{SO,USp} = {\rm Tr}\left[ QQ^\dagger e^{-V'} + 
 \lambda \left(e^{-V'{}^{\rm T}} J e^{-V'} - J \right) \right]\ ,
 \label{sbg:eq:so_usp_kahler}
\end{eqnarray}
where $Q$ are ${N_{\rm f}}$ chiral superfields as earlier and $V'$ is a vector
superfield of $U({N_{\rm c}})$. 
The added term breaks the complexified gauge transformations down to
$SO({N_{\rm c}}), USp(2{M_{\rm c}})$ and the equations of motion for
$\lambda$ gives the constraint (\ref{sbg:eq:V'}) which reduces the
K\"ahler potential (\ref{sbg:eq:so_usp_kahler}) back to
\eref{sbg:eq:Kahler_SOUSp}.  
The equation of motion for $V'$ takes the form
\begin{eqnarray}
QQ^\dagger e^{-V'} + \left(\lambda + \epsilon\lambda^{\rm T}\right)J =
0\ , \nonumber
\end{eqnarray}
where we have used (\ref{sbg:eq:V'}). $\lambda$ can be eliminated by
combining the equations of motion with their transposed ones. The
resultant equation contains the square of the manifest Hermitian
and positive definite matrix $X$ that traces to the K\"ahler potential
(\ref{sbg:eq:Kahler_SOUSp}) that we set out to find: 
\begin{align}
X^2 =  \left(Q^{\rm T} J \sqrt{QQ^\dagger}\right)^\dagger \left(Q^{\rm
  T} J \sqrt{QQ^\dagger}\right)\ , \quad
X \equiv \sqrt{QQ^\dagger}e^{-V'}\sqrt{QQ^\dagger}\ . \nonumber
\end{align}
We can uniquely obtain $V'$ from $X$ if and only if ${\rm
  rank}\, M>{N_{\rm c}}-2$, where $M$ are the holomorphic invariants for the
$SO, USp$ gauge theories, that is, the vacuum is in the full Higgs
phase.  
See Appendix B in Ref.~\refcite{Eto:2008qw} for a uniqueness proof, in
the case of ${\rm rank}\,M={N_{\rm c}}-1$.
It is possible to switch to $Q_{\rm wz}$ from $Q$ by the complexified
gauge transformation $Q_{\rm wz} = u'{}^{-1} Q$ with $u'u'{}^\dagger =
 e^{V'}$.
Without using an explicit solution for $V'$, we obtain the K\"ahler
potential of the NL$\sigma$M (according to method (\ref{sbg:method2}))
\begin{eqnarray}
K_{SO,USp} = {\rm Tr}\, X =
{\rm Tr} \sqrt{\left(Q^{\rm T} J \sqrt{QQ^\dagger}\right)^\dagger
  \left(Q^{\rm T} J \sqrt{QQ^\dagger}\right)}\ .
\label{sbg:eq:kahlerpot_so_sp_phi}
\end{eqnarray}

Now we can naturally switch to an expression according to method
(\ref{sbg:method3}) for this NL$\sigma$M. 
With the help of ${\rm Tr}_{\rm c}\sqrt{AA^\dagger} = 
{\rm Tr}_{\rm f}\sqrt{A^\dagger A}$, one can rewrite the K\"ahler
potential (\ref{sbg:eq:kahlerpot_so_sp_phi}) as follows
\begin{eqnarray}
K_{SO,USp}
= {\rm Tr}_{\rm f}\sqrt{M M^\dagger}\ ,\qquad
M^{\rm T} = \epsilon M\ ,
\label{sbg:eq:Kahler_SO,USp_N=1}
\end{eqnarray}
where $M$ is nothing but the holomorphic invariants of the gauge
symmetry
\begin{eqnarray}
M \equiv Q^{\rm T} J Q\ ,\qquad
B^{\left<A\right>} \equiv \det Q^{\left<A\right>}\ . \nonumber
\end{eqnarray}
The first one is the ``mesonic" invariant while the second is the
``baryonic" one which appears for ${N_{\rm f}} \ge {N_{\rm c}}$ and
$\langle A\rangle \equiv \langle A_1,\ldots,A_{N_{\rm c}}\rangle$ and 
$Q^{\langle A\rangle}$ is the $N_{\rm c}$-by-$N_{\rm c}$ minor matrix, 
i.e.~$\left(Q^{\langle A\rangle}\right)_{i,j} \equiv Q_{i,A_{j}}$ with
$A_j \in [1,N_{\rm f}]$.
The two kinds of invariants should be subject to constraints in order
to correctly express the NL$\sigma$M. 
There are relations between the mesons and the baryons:
\begin{align}
SO({N_{\rm c}}):\det(J) \ B^{\left<A\right>} B^{\left<B\right>} &=  
\det M^{\left<A\right>\left<B\right>} 
\label{sbg:eq:rel_MB_SO}\\
USp(2{M_{\rm c}}):\qquad\ \, {\rm Pf}(J)\ B^{\left<A\right>} &= 
{\rm Pf}\,M^{\left<A\right>\left<A\right>}. \nonumber
\end{align}
The Pl\"ucker relation among the baryonic invariants
$B^{\left<A\right>}$ is derived from the above relation.
Actually, from the invariants $M$ and $B^{\left<A\right>}$ with the
constraints 
we can reconstruct $Q$ modulo the complexified gauge symmetry as follows.
By using an algorithm similar to the Cholesky decomposition of an
Hermitian matrix, we show in Ref.~\refcite{Eto:2008qw} that
\begin{equation}\textrm{\parbox{0.8\linewidth}{
An arbitrary $n\times n$ (anti-)symmetric complex matrix $X$ can
always be decomposed as $X=p^{\rm T}Jp$ with $p$ being a 
${\rm rank}(X)\times n$ matrix.} }\end{equation}
In the $USp$ case, with a decomposition of the meson $M$, we can
completely reconstruct $Q$ modulo $USp(2{M_{\rm c}})^{\mathbb C}$ transformations.  
This fact corresponds to that there are no independent baryons
$B^{\left<A\right>}$ in a $USp(2{M_{\rm c}})$ theory and only the meson
fields describe the full Higgs phase
\begin{align}
{\cal M}_{USp}=\left\{M\,|\,M\in {N_{\rm f}} \times {N_{\rm f}} {~\rm matrix},\quad   
M^{\rm T}=-M,\quad {\rm rank}\,M=2{M_{\rm c}} \right\}\ . \nonumber
\end{align}
On the contrary, in the $SO({N_{\rm c}})$ case, a decomposition of $M$ gives
$Q$ modulo $O({N_{\rm c}})^{\mathbb C}$ and one finds two candidates for $Q$
since ${\mathbb Z}_2\simeq O^{\mathbb C}/SO^{\mathbb C}$ which is
fixed by the sign of the baryons.\footnote{ 
In the case of ${\rm rank}\,M={N_{\rm c}}-1$, 
$g\in {\mathbb Z}_2$ acts trivially on
$Q$ as $g\,Q=Q$, although all the baryons vanish.} 
Therefore we have to take the degrees of freedom of the baryons into
account to consider the full Higgs phase
\begin{align}
 {\cal M}_{SO}=
\Big\{M, B^{\left<A\right>}\,|&\,M:\,{\rm symmetric}~{N_{\rm f}}\times{N_{\rm f}},\nonumber\\
&{\rm Relation~(\ref{sbg:eq:rel_MB_SO})},\quad {N_{\rm c}}-1\le {\rm rank}\,M\le
{N_{\rm c}}\Big\}\ . \nonumber
\end{align}

For large ${N_{\rm c}}$, it is a hard task to obtain an explicit metric from
the formula (\ref{sbg:eq:Kahler_SO,USp_N=1}), since we need to calculate
the eigenvalues of $MM^\dagger$. In Ref.~\refcite{Eto:2008qw} we
calculate explicitly an expansion of the K\"ahler potential around
its vacuum value from which we are able to obtain the metric and
curvature.

\section{The $U(1) \times SO({N_{\rm c}})$ and $U(1)\times
  USp(2{M_{\rm c}})$ K\"ahler Quotients}

Next, we would like to consider a K\"ahler quotient by gauging an
overall $U(1)$ phase in addition to the $SO({N_{\rm c}})$ or $USp(2{M_{\rm c}})$ gauge
symmetry. 
We turn on the FI $D$-term associated with the additional $U(1)$ gauge
group. The K\"ahler potential can be written as
\begin{align}
K_{U(1)\times (SO,USp)} = {\rm Tr}\left[ QQ^\dagger e^{-V'} e^{-V_e} + 
\lambda \left(e^{-V'{}^{\rm T}} J e^{-V'} - J \right) \right] + \xi
V_e\ , \nonumber
\end{align}
where $V_e$ is the vector multiplet of the additional $U(1)$ gauge field.
We have already solved the $SO({N_{\rm c}})$ and $USp(2{M_{\rm c}})$ part in the
previous section, so the K\"ahler potential can be rewritten as
\begin{align}
K_{U(1)\times (SO,USp)} = {\rm Tr}\left[ \sqrt{ M M^\dagger}
  \right]e^{-V_e} + \xi V_e\ . \nonumber
\end{align}
The equation of motion for $V_e$ can be solved by
$V_e = \log\left[{{\rm Tr}\left(\sqrt{ M M^\dagger}\right)}/{\xi}\right]$.
Plugging this into the K\"ahler potential, we obtain
\begin{eqnarray}
K_{U(1)\times (SO,USp)} = \xi \log \left[{\rm Tr}\left(\sqrt{ M
    M^\dagger}\right)\right]\ ,\qquad
 M \equiv Q^{\rm T} J Q\ .
\label{sbg:eq:Kahler_pot_UxSO}
\end{eqnarray}

\section{The $SO({N_{\rm c}}),\ USp(2{M_{\rm c}})$ HyperK\"ahler Quotients}

Our next task is lifting up the $SO({N_{\rm c}})$ and 
$USp({N_{\rm c}}=2{M_{\rm c}})$ K\"ah{-}ler quotients to hyperK\"ahler
quotients. In order to construct the $SO({N_{\rm c}}),USp(2{M_{\rm c}})$
hyperK\"ahler quotients we need to consider ${\cal N}=2$
hypermultiplets. Hence, we consider an ${\cal N}=2$ extension of the
${\cal N}=1$ K\"ahler potential (\ref{sbg:eq:so_usp_kahler}), together
with the superpotential 
\begin{eqnarray}
\tilde K_{SO,USp} &=& {\rm Tr}\left[ QQ^\dagger e^{-V'} + \tilde Q^\dagger
  \tilde Q e^{V'} + \lambda \left(e^{-V'{}^{\rm T}} J e^{-V'} - J
  \right) \right]\ , \label{sbg:eq:hyper_kahler_so_usp_1}\\
W &=& {\rm Tr} \left[Q\tilde Q \Sigma' + \chi \left(\Sigma'{}^{\rm T}J + J 
  \Sigma'\right) \right]\ , \label{sbg:eq:superpot_so_usp_1}
\end{eqnarray}
where $(V',\Sigma')$ denote the $SO({N_{\rm c}})$ and $USp(2{M_{\rm c}})$ vector
multiplets, $(Q,\tilde Q)$ are ${N_{\rm f}}$ hypermultiplets in the
fundamental representation of $SO({N_{\rm c}})$ or $USp(2{M_{\rm c}})$, and
$(\lambda,\chi)$ are the Lagrange multipliers which are ${N_{\rm c}} \times
{N_{\rm c}}$ matrix valued superfields. We can rewrite the K\"ahler
potential (\ref{sbg:eq:hyper_kahler_so_usp_1}) as follows 
\begin{align}
\tilde K_{SO,USp}
= {\rm Tr} \left[ {\cal Q} {\cal Q}^\dagger e^{-V'}\right]\ , 
\qquad {\rm with} \qquad 
{\cal Q} \equiv \left( Q,\ J\tilde Q^{\rm T}\right)\ , \nonumber
\end{align}
where we have used that $e^{V'{}^{\rm T}} = J^{\rm T} e^{-V'} J$. This
trick relates the superfields in the anti-fundamental representation
with those of the fundamental representation via the algebra.
This K\"ahler potential is nothing but the ${\cal N}=1$ K\"ahler
potential of $SO({N_{\rm c}})$ and $USp(2{M_{\rm c}})$ with ${\cal Q}$
being a set of $2{N_{\rm f}}$ chiral superfields. 
We can straightforwardly borrow the result of
Sec.~\ref{sbg:sec:kahler_so_usp} and hence the K\"ahler potential reads 
\begin{eqnarray}
\tilde K_{SO,USp} = {\rm Tr}\left[\sqrt{{\cal M}{\cal M}^\dagger}\right]\ ,
\qquad {\cal M} \equiv {\cal Q}^{\rm T} J {\cal Q}\ .
\end{eqnarray}
The constraint coming from the superpotential
(\ref{sbg:eq:superpot_so_usp_1}) is 
\begin{eqnarray}
{\cal Q}\tilde J {\cal Q}^{\rm T}=0\ ,
\quad {\rm with~}
\tilde J\equiv 
\left(
\begin{array}{cc}
 {\bf 0}_{N_{\rm f}}& {\bf 1}_{{N_{\rm f}}} \\
-\epsilon {\bf 1}_{{N_{\rm f}}} & {\bf 0}_{N_{\rm f}}
\end{array}
\right)\ . \nonumber
\end{eqnarray}
Therefore, we again find the constraints for the meson field ${\cal M}$
\begin{eqnarray}
 {\cal M}^{\rm T}=\epsilon {\cal M}\ ,\quad {\cal M}\tilde J{\cal
   M}=0\ ,\quad
{N_{\rm c}}-2< {\rm rank}\,{\cal M}\le {N_{\rm c}}\ . \nonumber
\end{eqnarray} 
As is well-known, the $SO({N_{\rm c}})$ case has a $USp(2{N_{\rm f}})$
flavor symmetry while the $USp(2{M_{\rm c}})$ case has an $SO(2{N_{\rm
    f}})$ flavor symmetry.

Like in the case of the K\"ahler manifolds, the vacuum manifolds in
the hyperK\"ahler case can be written down for a generic point, which
for the $SO(N_{\rm c})$ case contains the space of
\eref{sbg:eq:SOvacuum_manifold} 
\begin{align}
\mathcal{M}_{SO(N_{\rm c})}^{\rm generic, HK} \simeq 
\mathbb{R}_{\geq 0}^{N_{\rm c}}\times \frac{USp(2N_{\rm f})}
{USp(2N_{\rm f}-2N_{\rm c})} 
\mathop\supset_{\begin{array}{c}\textrm{\tiny K\"ahler}\\[-7pt]\textrm{\tiny
      submfd.}\end{array}}
\mathcal{M}_{SO(N_{\rm c})}^{\rm generic, K}\ . \nonumber
\end{align}
Similarly, in a generic point on the vacuum manifold of the $USp(2M_{\rm
  c})$ hyperK\"ahler case, we can write
\begin{align}
\mathcal{M}_{USp(2M_{\rm c})}^{\rm generic, HK} \simeq  
\mathbb{R}_{\geq 0}^{M_{\rm c}}\times \frac{SO(2N_{\rm f})}
{SO(2N_{\rm f} - 4M_{\rm c})\times SU(2)^{M_{\rm c}}} 
\mathop\supset_{\begin{array}{c}\textrm{\tiny K\"ahler}\\[-7pt]\textrm{\tiny
      submfd.}\end{array}}
\mathcal{M}_{USp(2M_{\rm c})}^{\rm generic, K}\ , \nonumber
\end{align}
where \eref{sbg:eq:USpvacuum_manifold} is a special Lagrangian
sub-manifold of the 
hyperK\"ahler manifold and analogously for the $SO$ case.

Let us make a comment on the hyperK\"ahler quotient of the $U(1)\times
SO({N_{\rm c}})$ and $U(1)\times USp(2{M_{\rm c}})$ theories. 
We succeeded in constructing the hyperK\"ahler quotient of $SO({N_{\rm
    c}})$ and  $USp(2{M_{\rm c}})$ thanks to the fact that $J\tilde
Q^{\rm T}$ is in the anti-fundamental representation, which is the
same representation as $Q$. Although, we want to make use of the same
strategy for  $U(1) \times SO({N_{\rm c}})$ and $U(1) \times
USp(2{M_{\rm c}})$ as before, $J\tilde Q^{\rm T}$ still has charge
$-1$ with respect to the the $U(1)$ gauge symmetry while $Q$ has
$U(1)$ charge $+1$. Therefore, it is not an easy task to construct the
$U(1)\times SO({N_{\rm c}})$ and $U(1)\times USp(2{M_{\rm c}})$
hyperK\"ahler quotients.

\section{1/2 BPS states: NL$\sigma$M Lumps}

In this section, we will study the $\sigma$ model lumps which are 1/2
BPS states.  
Lumps are stringy topological defects in $d=1+3$ dimensional spacetime
and are supported by the non-trivial homotopy group $\pi_2({\cal M})$
associated with a holomorphic mapping from the 2 dimensional spatial
plane $z=x_1 + i x_2$ to the target space of the NL$\sigma$Ms
\cite{Polyakov:1975yp}.

Let us concentrate on $U(1)\times G'$ K\"ahler quotient models 
as NL$\sigma$Ms. 
In these cases, (inhomogeneous) complex coordinates on the K\"ahler
manifold  $\{\phi^\alpha\}$, which are the lowest scalar components of  
the chiral superfields, 
are given by some set of holomorphic $G'$ invariants;
$I^i$ modulo $U(1)^{\mathbb C}$, 
$\phi^\alpha \in \{I^i\}/\!\!/U(1)^{\mathbb C}$. 
Lump solutions can be obtained by just imposing $\phi^\alpha$ 
to be a holomorphic function
with respect to $z$ 
\begin{eqnarray}
\phi^\alpha(t,z,\bar z,x^3) \to \phi^\alpha (z;\varphi^i)\ ,
\label{sbg:eq:lump_sol}
\end{eqnarray}
where $\varphi^i$ denote complex constants. 
The mass of the lumps can be obtained by
plugging the solution back into the Lagrangian
\begin{eqnarray}
E = 2\int_{\mathbb{C}} \ K_{\alpha\bar \beta}(\phi,\bar \phi)
~\partial \phi^\alpha \bar \partial \bar \phi^{\bar \beta}\bigg|_{\phi \to
  \phi(z)}\ . \nonumber
\end{eqnarray}

We would like to stress that all the parameters $\varphi^i$ are
nothing but the moduli parameters of the 1/2 BPS lumps. 

We assume that the boundary i.e.~$z=\infty$ is mapped to 
a point $\phi^\alpha=\phi^\alpha_{\rm vev}$ 
on the vacuum manifold in a lump solution. 
Since the functions $\phi^\alpha(z)$ should be single valued,
$\phi^\alpha(z)$ can be expressed with a finite number of poles as  
\begin{eqnarray}
 \phi^\alpha(z)=\phi_{\rm
   vev}^\alpha+\sum_{i=1}^k\frac{\phi_i^\alpha}{z-z_i} 
+ \mathcal{O}(z^{-2})\ . \nonumber
\end{eqnarray}
Strictly speaking, we have to change patches of the manifold at the
poles in order to describe the solutions correctly.  To pick up such
global information of lumps thoroughly, it is convenient to use the
holomorphic invariants $I^i$ as homogeneous coordinates. 
By fixing some components of $I^i$ to be constants, we can construct
the invariants $I^i$ in terms of $\phi^\alpha(z)$ and find that $I^i$
also be holomorphic functions $I^i(z)=I_{\rm vev}^i+{\cal
  O}(z^{-1})$. We can redefine the functions $I^i(z)$ by using
$U(1)^{\mathbb C}$ transformations $I^i(z)\simeq
I^{'i}(z)=(z^\nu)^{n_i}I^i(z)$, such that all the invariants $I^i(z)$
are polynomials
\begin{eqnarray}
I^i(z)=I_{\rm vev}^i z^{n_i\nu }+{\cal O}(z^{n_i\nu-1})\ , \nonumber
\end{eqnarray}
where $n_i$ is the $U(1)$ charge of the holomorphic invariant $I^i$ 
and $n_i\,\nu\in {\mathbb Z}_{>0}$. These polynomials are basic tools 
to study lump solutions and their moduli, and $\phi^\alpha(z)$ can be
written as ratios of these polynomials, which are known as rational
maps in the Abelian case.
Here we assume that the invariants $I^i(z)$ have no common zeroes, 
in order to fix $U(1)^{\mathbb C}$, $I^i(z)\simeq (z-a) I^i(z)$. 
If a common zero accidentally emerges by varying the moduli parameters,
the behavior of lumps cannot be defined from the view point of the
NL$\sigma$M, since a common zero corresponds to the Coulomb phase for
$U(1)$ in the original gauge theory. This can also be understood as
the emergence of a local vortex (see Ref.~\refcite{Eto:2008qw} for
details).

Using the so-called {\it moduli matrix}, which describes different BPS
solitons in supersymmetric gauge theories \cite{Eto:2006pg}, we can
indeed identify the lowest component $\phi^\alpha$ with the moduli
matrix. The key observation is that the gauge symmetry $G$ in the
supersymmetric theory is naturally complexified: $G^{\mathbb
  C}$. Hence, the moduli matrix naturally appears in the superfield
formulation, while if we fix $G^{\mathbb C}$ in the Wess-Zumino gauge,
the scalar field $Q_{\rm wz}$ appears as the usual bosonic component
in the Lagrangian.

\section{Effective Action of Lumps}

Now we have a great advantage, thanks to the above superfield
formulation of the NL$\sigma$Ms. A supersymmetric low
energy effective theory of the 1/2 BPS lumps is immediately obtained
merely by plugging the 1/2 BPS solution (\ref{sbg:eq:lump_sol}) into the
K\"ahler potential which we have obtained in the previous section
after promoting the moduli parameters $\varphi^i$ to fields on the lump
world-volume 
\begin{eqnarray}
\phi^\alpha(t,z,\bar z,x^3) \to \phi^\alpha(z;\varphi^i(t,x^3))\ .
\nonumber
\end{eqnarray}
The resulting (effective) expression for the K\"ahler potential is
\begin{eqnarray}
{\cal K}_{\rm lump} = \int_{\mathbb{C}}  
K\left(\phi(z,\varphi^i(t,x^3),\ \phi^\dagger(\bar
z,\bar\varphi^i(t,x^3)\right)\ . \nonumber
\end{eqnarray}

\section{Lumps in $U(1)\times SO({N_{\rm c}})$ K\"ahler Quotients}

Let us start with a very simple example of a theory with the gauge
group $U(1) \times SO(2)$ and only two flavors ${N_{\rm f}}=2$. The
target space is\cite{Eto:2008qw} ${\mathbb C}P^1_+ \times {\mathbb
  C}P^1_-$. 
Lump solutions are classified by a pair of integers $(k_+,k_-)$ and
\begin{eqnarray}
\pi_2\left({\cal M}_{{N_{\rm f}}=2}^{U(1)\times SO(2)}\right) = {\mathbb Z}_+
\otimes {\mathbb Z}_-\ , \nonumber
\end{eqnarray}
where ${\mathbb Z}_{\pm}$ denote integers.
A solution with $(k_+,k_-)$ lumps is given by
\begin{eqnarray}
Q(z) = \begin{pmatrix}Q_{+1}(z) & Q_{+2}(z)\\Q_{-1}(z) & Q_{-2}(z)
\end{pmatrix} 
= \begin{pmatrix} \vec{Q}_+ \\ \vec{Q}_- \end{pmatrix} \ , \nonumber
\end{eqnarray}
where $Q_{\pm i}(z)$ are holomorphic functions of $z$ of
degree $k_{\pm}$, respectively. The energy reads
\begin{align}
E &=  2\int_{\mathbb{C}} \partial \bar\partial K_{U(1)\times SO(2)} =
\pi \xi (k_+ + k_-) \equiv \pi \xi k\ , \nonumber\\
K_{U(1)\times SO(2)} &= \frac{\xi}{2} \log |\vec Q_+|^2 +
\frac{\xi}{2} \log|\vec Q_-|^2 \ . \nonumber 
\end{align}
Interestingly, the tension of the minimal lump $(k_+,k_-)=(1,0),(0,1)$ 
is half of $2\pi\xi$ which is that of the minimal lump in the usual
${\mathbb C}P^1$ model.

We would now like to consider lump configurations in slightly
more complicated models by considering general $SO(2{M_{\rm c}})$ K\"ahler
quotients, where we set ${M_{\rm c}} \ge 2$, ${N_{\rm f}} = 2{M_{\rm c}}$ and $M_{\rm vev}=J$. 
However, we should take into account the following constraint on the
holomorphic invariants of the $SO(2{M_{\rm c}})$ group for $k$ lump
configurations \cite{Eto:2008yi} 
\begin{eqnarray}  M_{SO(2{M_{\rm c}})} = Q^{\rm T}(z) J Q(z) = J z^{k}
  + {\cal O}(z^{k-1})\ . 
\end{eqnarray}
As an example for $k=1$, we take
\begin{eqnarray}
Q_{k=1} = \begin{pmatrix}
z {\bf 1}_{M_{\rm c}} - {\bf A} & {\bf C}\\
0 & {\bf 1}_{M_{\rm c}}
\end{pmatrix} \ ,
\qquad
\left\{
\begin{array}{c}
{\bf A} = {\rm diag}(z_1,z_2,\cdots,z_{M_{\rm c}})\ ,\\
{\bf C} = {\rm diag}(c_1,c_2,\cdots,c_{M_{\rm c}})\ .
\end{array}
\right. \nonumber
\end{eqnarray}
This choice of diagonal matrices allows us to treat the invariants as
if they where simply invariants of $M$ different $SO(2)$ subgroups.
Note that non-zero parameters $c_i$ keep the ${\rm rank}\,M\ge 2{M_{\rm c}}$
even at $z=z_i$. Thus, the K\"ahler potential in
\eref{sbg:eq:Kahler_pot_UxSO} becomes
\begin{eqnarray} K = \xi \log \left( 2 \sum_{i=1}^M \sqrt{|z-z_i|^2 + |c_i|^2 }
\right)\ . \end{eqnarray}
The corresponding energy density is obtained by calculating
$\mathcal{E} = 2\partial\bar \partial K$ with this potential.
If we take some $c_i$ to vanish, the energy density becomes singular
at $z=z_i$ 
\begin{align}
 E=2\xi\partial\bar \partial\log\left[\sqrt{|z-z_i|^2}+\cdots\right]\sim 
{\rm const.}\times \frac{1}{|z-z_i|}+{\cal O}\left((z-z_i)^0\right)\ ,
\nonumber
\end{align} 
This is due to the occurrence of a curvature singularity in the
manifold when ${\rm rank}\,M=2{M_{\rm c}}-2$. 
The trace part of ${\bf C}$ determines the overall size of the
configuration and the trace part of ${\bf A}$ the center of mass,
where only the latter turns out to be a normalizable
mode\cite{Eto:2008qw}.

\section{Conclusion}

We have studied the NL$\sigma$M lumps in gauge theories with
$SO(N_{\rm c})$, $USp(2M_{\rm c})$, $U(1)\times SO(N_{\rm c})$ and 
$U(1)\times USp(2M_{\rm c})$ gauge groups and obtained K\"ahler metrics
and for the two first cases also the hyperK\"ahler
metrics. Furthermore, we have constructed NL$\sigma$M lumps in these
models.

\section*{Acknowledgments}
The author thanks the Organizers for warm hospitality.



\begin{thebibliography}{99}

\bibitem{Eto:2008qw}
  M.~Eto, T.~Fujimori, S.~B.~Gudnason, M.~Nitta and K.~Ohashi,
  arXiv:0809.2014 [hep-th].

\bibitem{AlvarezGaume:1981hm}
  L.~Alvarez-Gaume and D.~Z.~Freedman,
  Commun.\ Math.\ Phys.\  {\bf 80}, 443 (1981).

\bibitem{Curtright:1979yz}
  T.~L.~Curtright and D.~Z.~Freedman,
  Phys.\ Lett.\  B {\bf 90}, 71 (1980)
  [Erratum-ibid.\  B {\bf 91}, 487 (1980)]; 
  L.~Alvarez-Gaume and D.~Z.~Freedman,
  Phys.\ Lett.\  B {\bf 94}, 171 (1980);
  M.~Rocek and P.~K.~Townsend,
  Phys.\ Lett.\  B {\bf 96}, 72 (1980).

\bibitem{Lindstrom:1983rt}
  U.~Lindstr\"{o}m and M.~Ro\v{c}ek,
  Nucl.\ Phys.\  B {\bf 222}, 285 (1983).

\bibitem{Hitchin:1986ea}
  N.~J.~Hitchin, A.~Karlhede, U.~Lindstr\"{o}m and M.~Ro\v{c}ek,
  Commun.\ Math.\ Phys.\  {\bf 108}, 535 (1987).

\bibitem{Atiyah:1978ri}
  M.~F.~Atiyah, N.~J.~Hitchin, V.~G.~Drinfeld and Yu.~I.~Manin,
  Phys.\ Lett.\  A {\bf 65}, 185 (1978).

\bibitem{Kronheimer:1989zs}
  P.~B.~Kronheimer,
  J.\ Diff.\ Geom.\  {\bf 29}, 665 (1989).

\bibitem{Gibbons:1996nt}
  G.~W.~Gibbons, R.~Goto and P.~Rychenkova,
  Commun.\ Math.\ Phys.\  {\bf 186}, 585 (1997)
  [arXiv:hep-th/9608085].

\bibitem{Argyres:1996hc}
  P.~C.~Argyres, M.~Ronen Plesser and A.~D.~Shapere,
  Nucl.\ Phys.\  B {\bf 483}, 172 (1997)
  [arXiv:hep-th/9608129].

\bibitem{Seiberg:1994aj}
  N.~Seiberg and E.~Witten,
  Nucl.\ Phys.\  B {\bf 431}, 484 (1994)
  [arXiv:hep-th/9408099].

\bibitem{Argyres:1996eh}
  P.~C.~Argyres, M.~R.~Plesser and N.~Seiberg,
  Nucl.\ Phys.\  B {\bf 471}, 159 (1996)
  [arXiv:hep-th/9603042].

\bibitem{Antoniadis:1996ra}
  I.~Antoniadis and B.~Pioline,
  Int.\ J.\ Mod.\ Phys.\  A {\bf 12}, 4907 (1997)
  [arXiv:hep-th/9607058].

\bibitem{Eto:2005wf}
  M.~Eto, Y.~Isozumi, M.~Nitta, K.~Ohashi, K.~Ohta, N.~Sakai and Y.~Tachikawa,
  Phys.\ Rev.\  D {\bf 71}, 105009 (2005)
  [arXiv:hep-th/0503033].

\bibitem{Polyakov:1975yp}
  A.~M.~Polyakov and A.~A.~Belavin,
  JETP Lett.\  {\bf 22}, 245 (1975)
  [Pisma Zh.\ Eksp.\ Teor.\ Fiz.\  {\bf 22}, 503 (1975)].

\bibitem{Vachaspati:1991dz}
  T.~Vachaspati and A.~Achucarro,
  Phys.\ Rev.\  D {\bf 44}, 3067 (1991);
  A.~Achucarro and T.~Vachaspati,
  Phys.\ Rept.\  {\bf 327}, 347 (2000)
  [arXiv:hep-ph/9904229].

\bibitem{Eto:2007yv}
  M.~Eto {\it et al.},
  Phys.\ Rev.\  D {\bf 76}, 105002 (2007)
  [arXiv:0704.2218 [hep-th]].

\bibitem{Shifman:2006kd}
  M.~Shifman and A.~Yung,
  Phys.\ Rev.\  D {\bf 73}, 125012 (2006)
  [arXiv:hep-th/0603134].

\bibitem{Hanany:2003hp}
  A.~Hanany and D.~Tong,
  JHEP {\bf 0307}, 037 (2003)
  [arXiv:hep-th/0306150].

\bibitem{Luty:1995sd}
  M.~A.~Luty and W.~Taylor,
  Phys.\ Rev.\  D {\bf 53}, 3399 (1996)
  [arXiv:hep-th/9506098].

\bibitem{Bando:1983ab}
  M.~Bando, T.~Kuramoto, T.~Maskawa and S.~Uehara,
  Phys.\ Lett.\  B {\bf 138}, 94 (1984).

\bibitem{Intriligator:1995id}
  K.~A.~Intriligator and N.~Seiberg,
  Nucl.\ Phys.\  B {\bf 444}, 125 (1995)
  [arXiv:hep-th/9503179].

\bibitem{Intriligator:1995ne}
  K.~A.~Intriligator and P.~Pouliot,
  Phys.\ Lett.\  B {\bf 353}, 471 (1995)
  [arXiv:hep-th/9505006].

\bibitem{Eto:2006pg}
  M.~Eto, Y.~Isozumi, M.~Nitta, K.~Ohashi and N.~Sakai,
  J.\ Phys.\ A  {\bf 39}, R315 (2006)
  [arXiv:hep-th/0602170].

\bibitem{Eto:2008yi}
  M.~Eto, T.~Fujimori, S.~B.~Gudnason, K.~Konishi, M.~Nitta, K.~Ohashi and W.~Vinci,
  Phys.\ Lett.\  B {\bf 669}, 98 (2008)
  [arXiv:0802.1020 [hep-th]].

\end{thebibliography}
\end{document}